\documentclass[%
 amsmath,amssymb,
 longbibliography,
]{revtex4-1}

\usepackage{graphicx}
\usepackage{dcolumn}
\usepackage{bm}
\usepackage{hyperref}

\begin{document}


\title{Finite-size scaling in
globally coupled phase oscillators with a general coupling scheme}

\author{Isao Nishikawa$^{1}$, Koji Iwayama$^{1,2}$, Gouhei Tanaka$^{1,3}$, Takehiko Horita$^{4}$,
 and Kazuyuki Aihara$^{1,3}$}
\affiliation{
$^{1}$Institute of Industrial Science, The University of Tokyo, Tokyo 153-8505, Japan
}
\affiliation{
$^{2}$FIRST, Aihara Innovative Mathematical Modelling Project,
JST, Tokyo 153-8505, Japan
}
\affiliation{
$^{3}$Department of Mathematical Informatics, Graduate School of Information Science and Technology, The University of Tokyo, 7-3-1 Hongo, Bunkyo-ku, Tokyo 113-8656, Japan
}
\affiliation{
$^{4}$Department of Mathematical Sciences, Osaka Prefecture University,
Sakai 599-8531, Japan
}




\date{\today}

\begin{abstract}

We investigate a critical exponent related to synchronization transition in globally coupled nonidentical phase oscillators.
The critical exponents of susceptibility, correlation time, and correlation size are significant quantities
to characterize fluctuations in coupled oscillator systems of large but finite size and understand a universal property of synchronization.
These exponents have been identified for the sinusoidal coupling but not fully studied for other coupling schemes.
Herein, for a general coupling function including a negative second harmonic term in addition to the sinusoidal term,
we numerically estimate the critical exponent of the correlation size, denoted by $\nu _+$,
in a synchronized regime of the system by employing a non-conventional statistical quantity.
First, we confirm that the estimated value of $\nu _+$ is approximately 5/2 for the sinusoidal coupling case,
which is consistent with the well-known theoretical result.
Second, we show that the value of $\nu _+$ increases with an increase in the strength of the second harmonic term.
Our result implies that the critical exponent characterizing synchronization transition largely depends on the coupling function.

\begin{description}
\item[PACS numbers]
64.60.F-, 05.45.Xt

\end{description}
\end{abstract}

\pacs{Valid PACS appear here}
\maketitle



\section{Introduction}
\begin{figure}[h]
 \includegraphics[width=6cm]{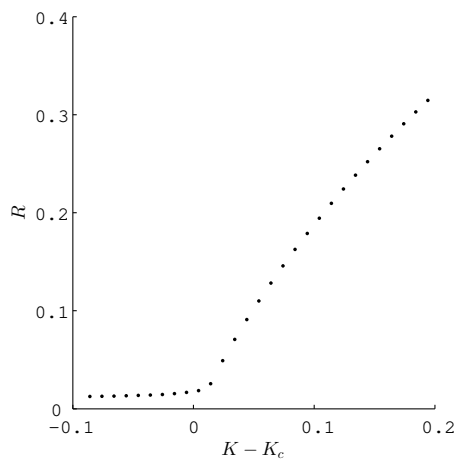}
 \caption{Bifurcation diagram of the order parameter $R$, where $h(x) = \sin (x) - 0.5 \sin(2x)$ and $N=8000$. A supercritical bifurcation occurs.}
 \label{fig:k-r}
\end{figure}

Populations of coupled rhythmic elements can exhibit synchronization and collective behavior via mutual interactions \cite{PikovskyBook}.
Such phenomena are observed in a variety of systems such as chemical reactions, engineering circuits, and biological populations \cite{PikovskyBook}.
To elucidate the general properties of such phenomena,
the phase description of systems has been widely used \cite{PikovskyBook,KuramotoBook}.
In particular, there have been many studies on
globally coupled phase oscillators \cite{KuramotoBook,Kuramotomodel}, defined as follows:
\begin{align}
\label{phasemodel}
\dot \theta_j = \omega_j + \frac{K}{N} \sum _{k=1} ^N h(\theta_k - \theta _j), \ j=1,\cdots,N,
\end{align}
where $\theta _j$ is the phase of the $j$th oscillator,
$\omega_j$ is the natural frequency of the $j$th oscillator,
$K>0$ is the coupling strength,
$h$ is the coupling function, and $N$ is the number of oscillators.
When $h(x) = \sin(x)$, this model is referred to as the Kuramoto model \cite{KuramotoBook}.
One of the main issues in this model has been the scaling property of the order parameter defined as follows \cite{KuramotoBook}:
\begin{align}
\label{R}
R(t) \equiv \frac {1}{N} \left|\sum _{j=1} ^N \exp(2\pi i\theta _j)\right|.
\end{align}
In the thermodynamic limit ($N \rightarrow \infty $),
the phase oscillator model (\ref{phasemodel}) exhibits a synchronization transition when the coupling strength $K$ surpasses a critical value $K_c$.
This transition can be characterized by a change of the order parameter from zero to a non-zero value.
We assume that the stationary state $(R(t)=0$) in the incoherent regime supercritically bifurcates at the critical coupling strength $K=K_c$,
above which the oscillators are synchronized.
The behavior of the order parameter is exemplified for a finite-size system in Fig.~$\ref{fig:k-r}$ \cite{PikovskyBook,KuramotoBook,Kuramotomodel,Chiba,Chiba2,Chiba3}.
The scaling law of the order parameter with respect to a change in the coupling strength $K$ around the synchronization transition point
has been well studied in relation to the second order phase transition \cite{KuramotoBook,Sakaguchi2,Daido2,Crawford2,Chiba2,Chiba}.
The scaling property has been fully understood in the case where $N$ is infinite,
not only for the sinusoidal coupling function but also for general coupling functions \cite{Daido2,Crawford2,Chiba2}.
However, it is less clear in finite-size systems because the property of fluctuations depends on the system size.

The critical exponents of the order parameter, correlation time, susceptibility (corresponding to the product of the variance of the order parameter and $N$),
and correlation size (representing the number of oscillators which are almost synchronized but not completely)
are significant quantities
that can characterize the behavior near phase transitions in physical systems.
The values of the critical exponents of these statistical quantities can be used to categorize physical systems into minor classes,
because the values are independent of the details of the system \cite{Nishimori}.
Further, in equilibrium systems,
the critical exponents of these statistical quantities completely determine
those of all the other statistical quantities \cite{Nishimori}.
Therefore, the evaluation of the critical exponents in the phase oscillator model (\ref{phasemodel})
enables the clarification of differences between equilibrium and non-equilibrium systems.
In particular,
fluctuations in the systems of large but finite size can be characterized by
the exponents of susceptibility, correlation time, and correlation size.
Although the critical exponents
in the phase oscillator model (\ref{phasemodel}) with finite large $N$
have been obtained for the sinusoidal coupling function \cite{Daido,Pikovsky4,Hong,Hildebrand,Buice,Hong3},
it is known that the critical exponent of the order parameter depends on the coupling scheme \cite{Daido2,Crawford2}.
This fact motivated us to examine if the critical exponents of other statistical quantities also depend on the coupling function or not.

In the present paper,
we employ a non-conventional statistical quantity to evaluate the critical exponent of correlation size, $\nu _+$, in the synchronized regime of the phase oscillator model (\ref{phasemodel}) with finite large $N$.
This is because it is difficult to compute the value of $\nu_+$ using the critical exponent of the order parameter \cite{Hong3}.
The statistical quantity that we use is denoted by $D$, which is given by the diffusion coefficient of the temporal integration of the order parameter,
multiplied by system size $N$ \cite{Nishikawa}.
Using the statistical quantity $D$, we perform the finite-size scaling analysis \cite{Nishimori}.
First, we confirm that the estimated value of $\nu _+$ is approximately 5/2 for the sinusoidal coupling $h(x) = \sin (x) $,
which is consistent with the well-known theoretical result \cite{Hong,Hong2}.
Second, we consider a general coupling function including a negative second harmonic term in addition to the sinusoidal term, i.e. $h(x) = \sin (x) - q \sin(2x)$ with $q>0$.
We show that the value of $\nu _+$ increases with an increase in the strength $q$ of the second harmonic term.
Our result means that the critical exponent characterizing synchronization transition largely depends on the coupling function.

Although fluctuations of the order parameter in the phase oscillator model (\ref{phasemodel}) have been studied
for the past two decades \cite{Daido,Pikovsky4,Hong,Hong3,Hildebrand,Buice,Nishikawa},
those for a general coupling function have not been fully understood.
In particular, for any general coupling scheme other than the sinusoidal one,
the critical exponents of statistical quantities have not been reported
except for that of the non-conventional statistical quantity $D$ \cite{Nishikawa}.
Note that the values of critical exponents are independent of the details of systems and they characterize universal structures of the systems \cite{Nishimori}.
In fact, in the thermodynamic limit $N \rightarrow \infty $, it was previously shown that as far as the coupling function includes a second harmonic term,
the critical exponent of the order parameter takes the same value \cite{Daido2,Crawford2}.
In contrast, our finding means that the value of $\nu _+$ crucially depends on the strength of the second harmonic term.
Our result highlights a new relation between the couping schemes and universal structures in the globally coupled phase oscillators (\ref{phasemodel}).

\section{Statistical quantity for the finite-size scaling analysis}

\begin{figure}[h]
 \includegraphics[width=6cm]{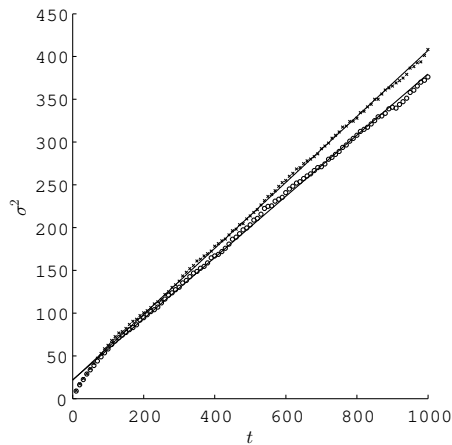}
 \caption{The time evolutions of the variance $\sigma ^2 (t)$ of the integrated order parameter
in the coherent regime of the Kuramoto model where $h(x) = \sin(x)$, $N=1000$, and
$K=1.635 > K_c = 1.59\cdots$.
These two evolutions are generated for different initial conditions.
The value of $\sigma ^2(t)$ increases linearly with slope $2Dt$ after a transient period in both cases.
Because of multistability,
the value of $D$ differs depending on the initial conditions.}
 \label{fig:estimation}
\end{figure}

\begin{figure}[h]
 \includegraphics[width=6cm]{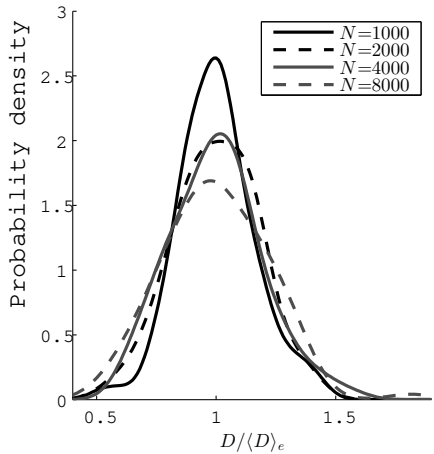}
 \caption{Probability density function of $D$ estimated by using 300 different initial conditions in the model (\ref{phasemodel}),
where $h(x) = \sin (x)$ and $K=1.635 > K_c = 1.59\cdots$.
The horizontal axis is normalized by $\langle D \rangle _e$,
where $\langle \cdot \rangle _e$ is the ensemble average over 300 different initial conditions.
We numerically obtained 300 values of $D$ for each $N$ and
estimated the probability density function of $D$ by means of kernel density estimation.
The standard deviation of $D$ is sufficiently small for each $N$.}
 \label{fig:disD}
\end{figure}

We consider the diffusion coefficient of the time integral of $R(t)$, which characterizes long-term fluctuations in $R(t)$ \cite{Nishikawa}.
The variance of $\int _0 ^t R(s) ds$ is given by:
\begin{align}
\label{sigma}
\displaystyle \sigma^2(t) \equiv N \lim _{T \rightarrow  \infty } \frac{1}{T} \int _{t_0 =0} ^{t_0 = T} \left( \int _{s=t_0} ^{s=t+t_0} R(s){\rm d}s - \langle R \rangle _t  t \right) ^2  {\rm d}{t_0},
\end{align}
where $\langle R \rangle _t$ represents the time average,
defined as follows:
\begin{align}
\label{average1}
\langle R \rangle _t \equiv \lim _{T \rightarrow  \infty } \frac{1}{T} \int _0 ^T R (t) {\rm d}t.
\end{align}
The following diffusion law then holds:
\begin{align}
\label{D}
D = \lim _{t\rightarrow \infty } \sigma^2(t) /  2t.
\end{align}
The value of $D$ characterizes how $\int _0 ^t R(s)ds$ differs from its mean value $\langle R \rangle _t t$ on the ensemble average.
The value of $D$ is estimated by fitting $\sigma ^2(t)$ with a line of slope $2Dt$ except for a transient period as shown in Fig.~$\ref{fig:estimation}$.
Different initial conditions may yield different values of $D$ because of multistability \cite{Maistrenko}, as seen in Fig.~$\ref{fig:estimation}$.
However, the dependence of $D$ on initial conditions is sufficiently small as shown in Fig.~$\ref{fig:disD}$.
This result implies that the property on fluctuations is almost similar for the multiple coexisting solutions.
All of them simultaneously transit from a non-synchronous state to a synchronous one as the coupling strength $K$ exceeds the critical value $K=K_c$
for a large system size $N$ \cite{PikovskyBook,KuramotoBook,Kuramotomodel,Chiba,Chiba2,Chiba3,Daido3}.




\begin{figure}[h]
\begin{center}
 \includegraphics[width=6cm]{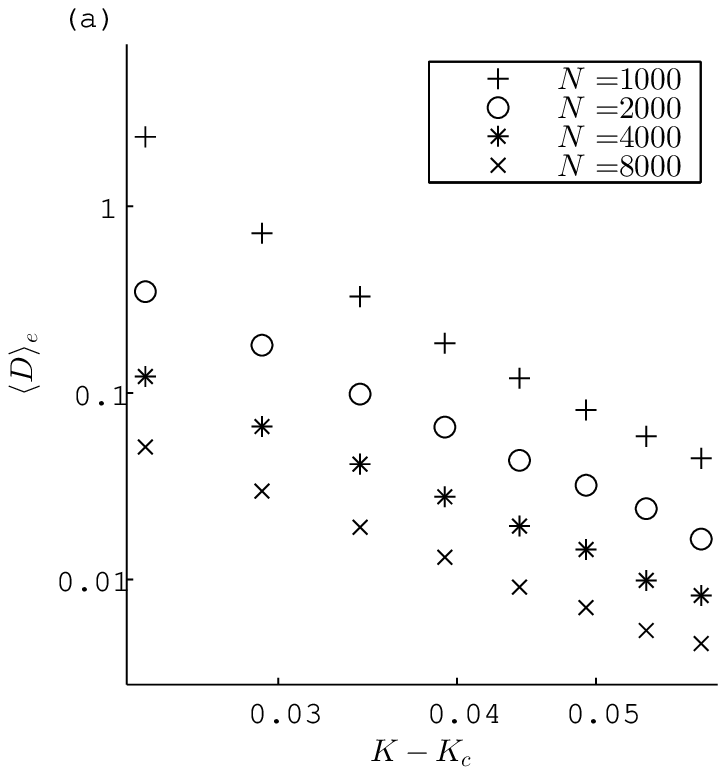}
 \includegraphics[width=6cm]{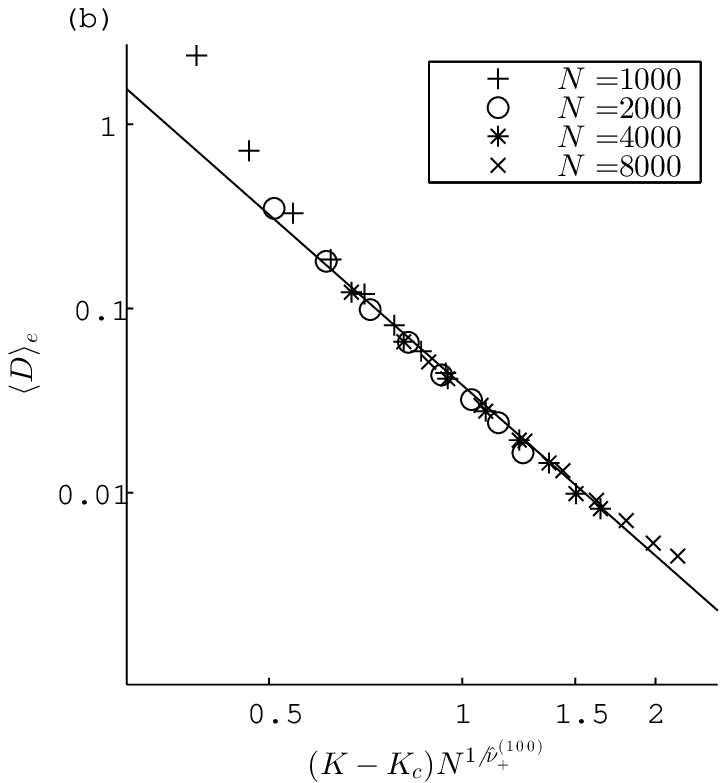}
\end{center}
 \caption{Estimation of $\nu _+$ in the Kuramoto model.
(a) $(K-K_c)$ vs. $\langle D \rangle _e$, where $\langle \cdot \rangle _e$ means the ensemble average over 100 overlapping samples.
(b) $(K-K_c)N^{1/2.49}$ vs. $\langle D \rangle _e$ in a log-log plot, which implies $\hat {\nu}_+ ^{(100)} \sim 2.49$.
The two left endpoints for $N=1000$ were not used for the line fitting.}
 \label{kuramoto_fit}
\end{figure}


\begin{figure}
 \includegraphics[width=6cm]{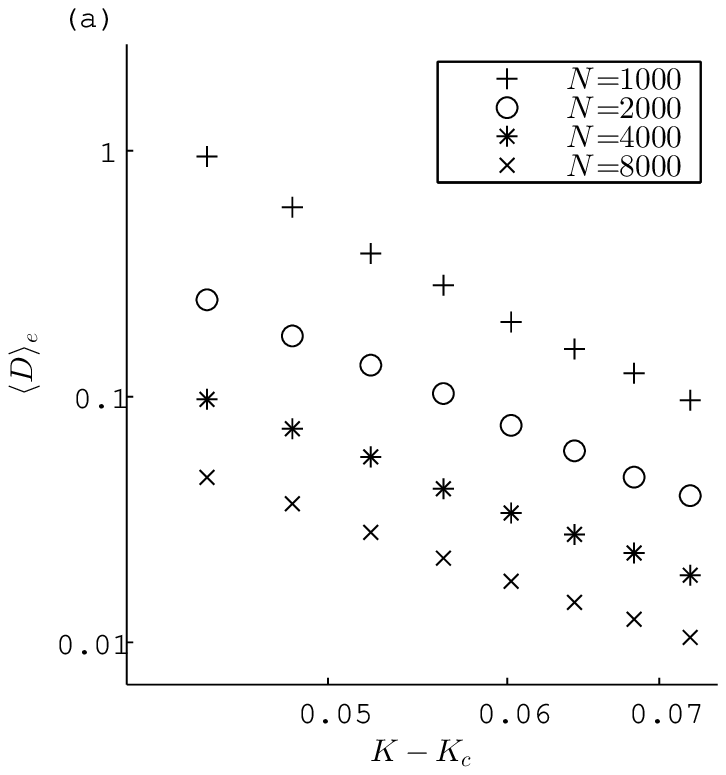}
 \includegraphics[width=6cm]{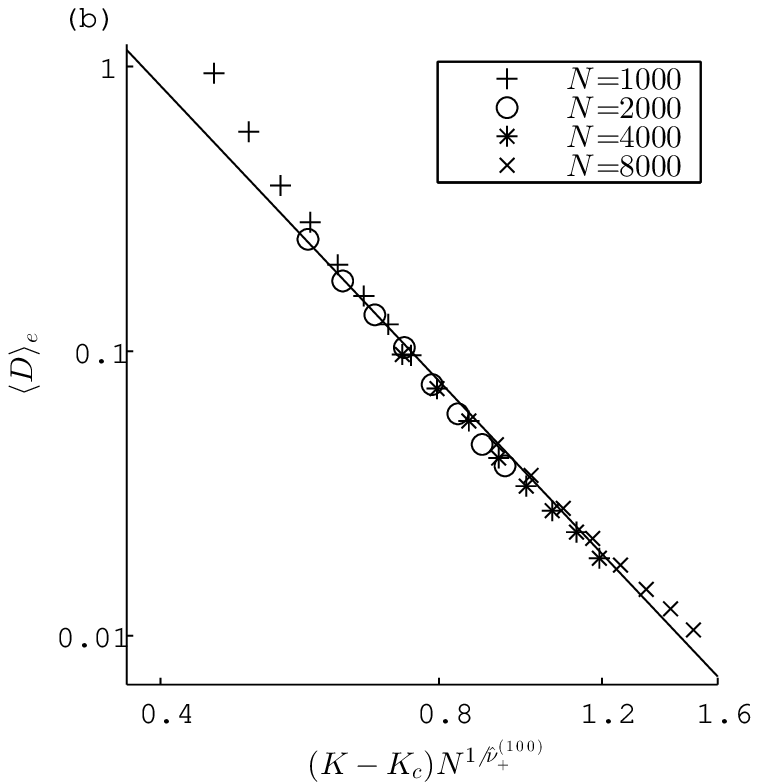}
  \caption{Estimation of $\nu _+$ in the phase oscillator model (\ref{phasemodel}) with $h(x)=\sin(x) - 0.1 \sin(2x)$.
(a) $(K-K_c)$ vs. $\langle D \rangle _e$, where $\langle \cdot \rangle _e$ means the ensemble average over 100 overlapping samples.
(b) $(K-K_c)N^{1/2.95}$ vs. $\langle D \rangle _e$ in a log-log plot, which implies $\hat {\nu}_+ ^{(100)} \sim 2.95$.
The two left endpoints for $N=1000$ were not used for the line fitting.}
 \label{daido_fit}
\end{figure}

\section{The finite-size scaling analysis}
This study addresses the coupling function in the form $h(x) = \sin(x) - q \sin(2x) $
with $0\leq  q \leq  0.1$.
We examine this coupling function
because it is considered to be a ``generic" coupling function \cite{Daido2,Crawford2};
provided that the coupling function includes the second harmonic term in addition to the sinusoidal one,
the critical exponent of the order parameter does not depend on other higher-order terms \cite{Daido2,Crawford2}.

The scaling hypothesis \cite{Nishimori} states that any quantity $A$,
which shows critical divergence at $K=K_c$, is scaled for $K>K_c$ as follows:
\begin{align}
\label{scalinglaw2}
N^{-r_+/\nu _+} A = \Psi((K-K_c)N^{1/\nu _+}),
\end{align}
where $\nu _+$ and $r_+$ represent the critical exponents of the correlation size and $A$ in the synchronized regime, respectively.
The function $\Psi $ is approximated as
$\Psi(x) \sim ux^{-r_+}+x^{-s}$ with $s>r_+$ and a constant $u$ for large $x$ \cite{Nishimori}
so that the orders with respect to system size $N$ in both sides of Eq.~(\ref{scalinglaw2}) are consistent.
It was shown that for a general coupling function $h(x)$ in the phase oscillator model ($\ref{phasemodel}$),
the critical exponent of $D$ in the synchronized regime is equal to $0$ \cite{Nishikawa}.
Therefore, we replace $A$ by $D$ and then substitute $r_+=0$ into Eq.~($\ref{scalinglaw2}$).
Furthermore, we assume $u=0$ in the above approximation form of $\Psi(x) $ because $\Psi(x) \rightarrow 0$ as $x\rightarrow \infty $.
As a result, we obtain
\begin{align}
\label{scalinglaw}
D \sim ((K-K_c)N^{1/\nu _+})^{-s},
\end{align}
which means that
if the finite-size effect is not so strong,
(i) $D$ decreases in a power law fashion with system size $N$ for a fixed value of $K$,
and
(ii) the numerically computed values of $D$ should fall on a straight line against $(K-K_c)N^{1/\nu_+}$ in a log-log plot.

In numerical simulations,
the natural frequencies $\omega _j$ of the individual oscillators are chosen to satisfy
\begin{align}
\label{deterministic}
j/(N+1) = \int _{- \infty } ^{\omega _j} g(\tilde \omega ) {\rm d}\tilde \omega ,
\end{align}
where $g(\tilde \omega )$ is the Gaussian distribution with mean zero and variance one.
To avoid a situation in which the finite-size effect yields an erroneous value for $\nu _+$,
we limit the range of $K$ to $(K_c <) K_- < K < K_+$.
We select the lower value $K_-$ such that $D$ decreases in a power law fashion with system size $N$ for $K>K_-$.
Furthermore, we choose the upper value $K_+$ such that the power law decay of $D$
is kept within the range $K_- <K< K_+$
in order to avoid a large variation in the estimated values of $\nu _+$.
In addition, we use a bootstrap method.
First, we calculate the value of $D$ by using 100 different initial conditions for each pair of $(N, K)$.
We randomly choose $M$ overlapping samples of $D$ values from the 100 simulation results and average them.
Next,
we estimate the value of $\hat{\nu}_ + ^{(M)}$ that minimizes the mean squared
error between the mean values of $D$ and the fitting line in a log-log plot,
where $\hat{\nu} _+ ^{(M)}$ represents the estimated value of  $\nu _+$ using the averaged values of $D$ over the $M$ samples.
By repeating this procedure $10000$ times,
we obtain the estimated values of the mean and the standard deviation of $\hat{\nu} _+ ^{(M)}$.

\begin{figure}
 \includegraphics[width=6cm]{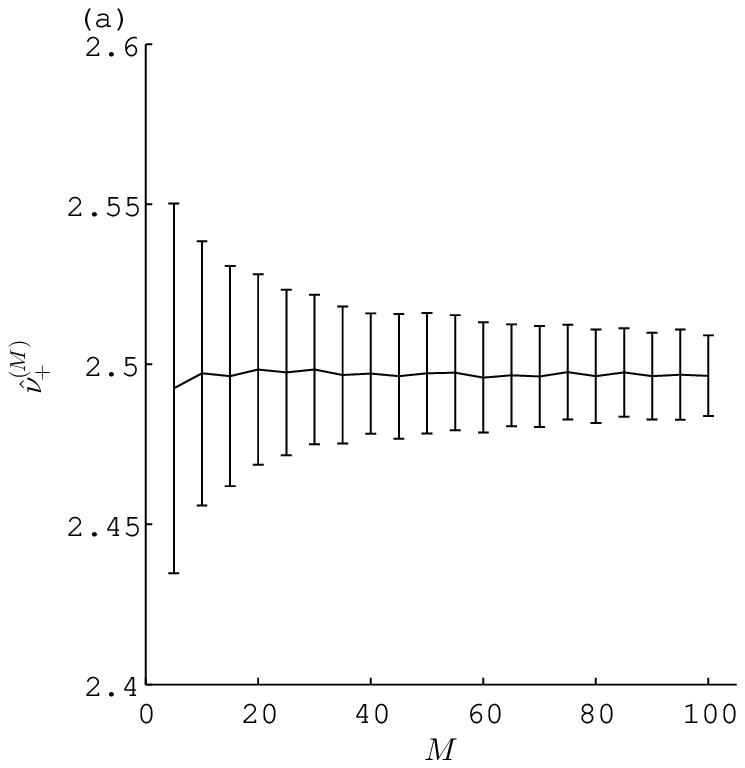}
 \includegraphics[width=6cm]{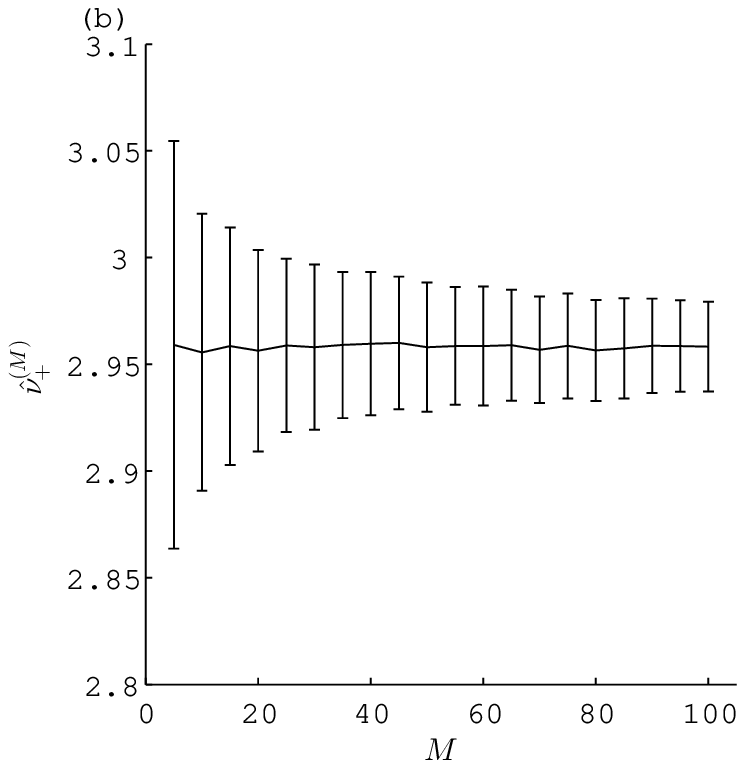}
 \caption{The standard deviation of the values of $\hat{\nu} _+ ^{(M)}$ estimated from $M$ samples, where $5 \leq M \leq 100$.
 (a) The case of $q=0$. (b) The case of $q=0.1$.
 In both cases, the standard deviation is very small for a sufficiently large $M$.}
 \label{fig:nu_bootstrap}
\end{figure}

\begin{figure}
 \includegraphics[width=6cm]{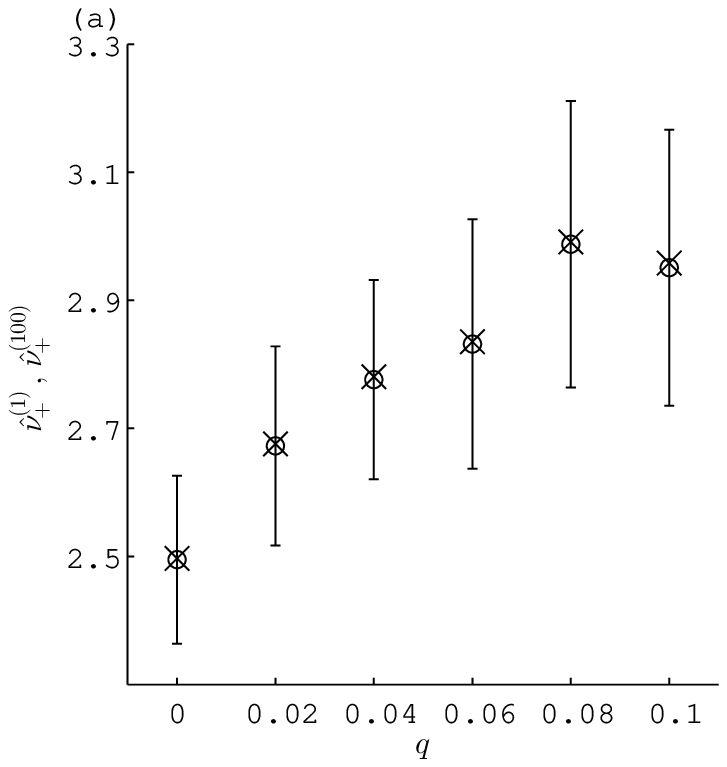}
 \includegraphics[width=6cm]{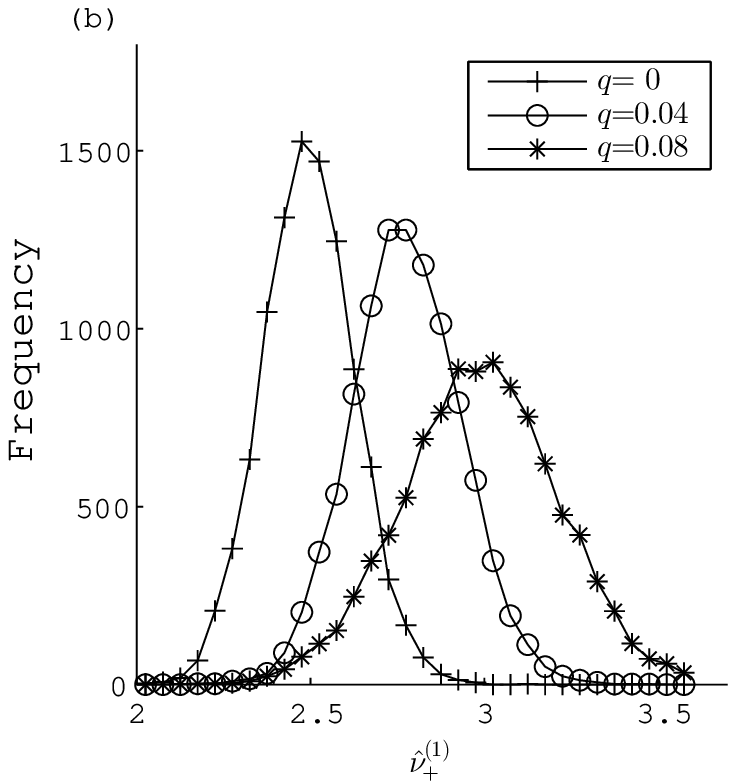}
 \caption{Relationship between the estimated exponent $\hat{\nu} _+ ^{(M)}$ and the strength $q$ of the negative second harmonic term
in the phase oscillator model ($\ref{phasemodel}$) with $h(x) = \sin (x) - q \sin (2x)$.
(a) The crosses indicate the mean values of $\hat {\nu }_+ ^{(100)}$. 
The circles and the error bars indicate the averages and the standard deviations of the distributions of $\hat{\nu}_+ ^{(1)}$, respectively.
(b) Distributions of $\hat{\nu}_+ ^{(1)}$ for each $q$.
}
 \label{nu_dist}
\end{figure}

First, let us consider the Kuramoto model, i.e. $q=0$.
The estimated mean value of $\hat{\nu}_ + ^{(100)}$ is given as $\hat{\nu}_ + ^{(100)} \sim  2.49$ by using the method explained above.
Figure~$\ref{kuramoto_fit}$(a) shows the value of $D$ averaged over 100 overlapping samples for each pair of $(N, K)$.
The same data points shown in Fig.~$\ref{kuramoto_fit}$(b) are plotted against $(K-K_c)N^{1/2.49}$ in a log-log plot.
The data are fitted well by the straight line.
The estimated exponent $\hat{\nu} _+ ^{(100)} \sim  2.49$ is consistent with the analytical results $\nu _+ = 5/2$ \cite{Hong,Hong2}.
Therefore, our method can estimate the exact value of $\nu _+$ well.


Next, we investigate the model ($\ref{phasemodel}$) with a more general coupling function with $q=0.1$.
Figure $\ref{daido_fit}$(a) shows the averaged values of $D$ over 100 overlapping samples.
The estimated mean value of $\hat{\nu}_ + ^{(100)}$ is given as $\hat{\nu}_ + ^{(100)} \sim  2.95$, as seen in Fig.~$\ref{daido_fit}$(b).
This means that the value of $\nu_+$ in this model is 
larger than that in the Kuramoto model.

Note that $M=100$ samples are enough to estimate the accurate value of $\nu _+$
because
the standard deviation of the estimated values of $\hat{\nu}_ + ^{(100)}$ is sufficiently small, as shown in Figs~$\ref{fig:nu_bootstrap}$(a) and $\ref{fig:nu_bootstrap}$(b).

Finally, we examine the relationship between the exponent $\nu _+$ and the strength $q$ of the negative second harmonic term.
The mean of $\hat{\nu} _+ ^ {(100)}$ is likely to increase with an increase in the value of $q$ as shown in Fig.~$\ref{nu_dist}$(a).
In addition, we obtain the distributions of $\hat{\nu} _+ ^ {(1)}$ as shown in Fig.~$\ref{nu_dist}$(b).
The standard deviation of $\hat{\nu} _+ ^ {(1)}$ is not so large and
the mean value of $\hat{\nu} _+ ^ {(1)}$ is also likely to increase with an increase in the value of $q$.
These results imply that the critical exponent $\nu _+$
depends on the second harmonic term of the coupling function.

\begin{figure}
 \includegraphics[width=6cm]{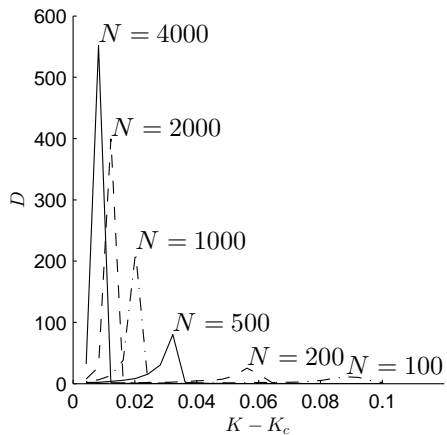}
  \caption{Relationship between the diffusion coefficient $D$ and the system size $N$.
$D$ takes its maximal value in the coherent regime due to the finite-size effect.
The value of the coupling strength $K$ exhibiting the maximal value of $D$ gradually approaches $K=K_c$ as $N$ increases.
A fixed initial condition is used to compute the value of $D$ for each $N$.}
 \label{fig:plot_KD}
\end{figure}

\section{Discussion}

We explain why our method yields a good estimation for the critical exponent $\nu _+$.
In the limit $N  \rightarrow  \infty $,  $D  \rightarrow  \infty $ as  $K  \rightarrow  K_c - 0 $,
whereas $D=0$ for $K>K_c$ \cite{Nishikawa}.
However, for a finite $N$, $D$ takes its maximal value in the coherent regime
as shown in Fig.~$\ref{fig:plot_KD}$.
It means that $D$ virtually reflects the feature of the incoherent solution of the infinite-size system even for $K>K_c$
if $K\simeq K_c$ and $N$ is finite.
Such a finite-size effect is well known in the literature \cite{Nishimori}.
As a result, to well estimate the value of $\nu _+$  by the finite-size scaling analysis,
we must avoid the coherent regime that is very close to the transition point $K=K_c$.
Our numerical simulations have selected the region in which the finite-size effect is weak enough so that $D\sim O(1/N^a)$.
Note that it is difficult to determine whether the finite-size effect is sufficiently weak by using other statistical quantities.

We remark on another numerical study about the critical exponent of correlation size \cite{Hong3}.
Let us denote the critical exponent of the order parameter $R$ by $\beta $ and that of correlation size in the incoherent regime
by $\nu _-$.
It is known that $R \sim N^{- \beta/\nu }$ for $K = K_c$ if $\nu _+ = \nu _- (=: \nu )$.
The value of $\nu $ can be computed as
$\nu = 5/4$ for the Kuramoto model with deterministically chosen natural frequencies \cite{Hong3}.
However, there is little evidence for $\nu _+ = \nu _-$.
In fact, as we have already mentioned,
the value of the critical exponent of $D$ differs depending on whether the system behavior is coherent or incoherent \cite{Nishikawa}.

Our numerical simulations have shown that
when the coupling function possesses a negative second harmonic term with strength $q$ in addition to the sinusoidal one,
$\nu _+ > 5/2$ and $\nu _+$ increases with $q$.
This result is consistent with the following property.
Near the synchronization transition point $K=K_c$, $R$ is scaled as
$R \sim (1+1/q)(K - K_c) + O((K-K_c)^2)$ \cite{Chiba2}.
This implies that the more the value of $q$ increases,
the more slowly the fluctuations of this system decay with the coupling strength $K$.

Our method has the following potential application.
The method in this paper enables us to obtain an approximate value of $\nu _+$ for a general coupling function.
The value of $\nu _+$ is the same for finite-size scaling analysis of other statistical quantities \cite{Nishimori}.
Suppose that we analytically obtain the value of $\nu _+$ or the critical exponents of other statistical quantities
such as susceptibility and correlation time for a general coupling function;
our methods make it possible to confirm the validity of the analytical results numerically by the finite-size scaling analysis \cite{Nishimori}
because we can compute an approximate value of $\nu _+$.

\section*{Acknowledgments}

This research is supported by
Grant-in-Aid for Scientific Research (A) (20246026) from MEXT of Japan,
and by
the Aihara Innovative Mathematical
Modelling Project, the Japan Society for the Promotion of Science
(JSPS) through the ``Funding Program for World-Leading Innovative R\&D
on Science and Technology (FIRST Program)," initiated by the Council
for Science and Technology Policy (CSTP).

\end{document}